# Diversity Extraction for Multicarrier Continuous-Variable Quantum Key Distribution

Laszlo Gyongyosi
[1] Quantum Technologies Laboratory
Department of Networked Systems and Services
Budapest University of Technology and Economics
2 Magyar tudosok krt., Budapest, H-1117 Hungary
[2] MTA-BME Information Systems Research Group
Hungarian Academy of Sciences
7 Nador st., Budapest, H-1051 Hungary

*Abstract*— We introduce a diversity extraction for multicarrier continuous-variable (CV) quantum key distribution (QKD). The diversity extraction utilizes the resources that are injected into the transmission by the additional degrees of freedom of the multicarrier modulation. The multicarrier scheme granulates the information into Gaussian subcarrier CVs and divides the physical link into several Gaussian sub-channels for the transmission. We prove that the exploitable extra degree of freedom in a multicarrier CVQKD scenario significantly extends the possibilities of single-carrier CVQKD. The diversity extraction allows for the parties to reach decreased error probabilities by utilizing those extra resources of a multicarrier transmission that are not available in a single-carrier CVQKD setting. The additional resources of multicarrier CVQKD allow the achievement of significant performance improvements that are particularly crucial in an experimental scenario.

## I. Introduction

By utilizing the fundamental laws of quantum mechanics, the continuous-variable quantum key distribution (CVQKD) systems allow to realize an unconditionally secure communication through the currently established communication networks. The CVQKD protocols do not require single photon devices in contrast to the first developed discrete-variable (DV) QKD protocols [1–17]. This significant benefit has immediately made possible to achieve the practical implementation of QKD by the standard devices of traditional telecommunications [18–19], [23–30]. In a CVQKD setting, the information is carried by Gaussian random distributed position and momentum quadratures, which identify a quantum state in the phase space. The quantum states are sent through a noisy link (e.g., an optical fiber or a wireless optical channel [18–19], [26-30]), which adds a white Gaussian noise to the phase space transmission. Despite the fact that the noise characteristic of a CVQKD transmission is plausible and well exploitable in the security proofs, the performance of CVQKD, particularly the currently available secret key rates, is still below the rates of the protocols of traditional telecommunications. This issue brings up a potential requirement on the delivery of an intensive performance enhancement for CVQKD. In particular, for this purpose the multicarrier CVQKD modulation has been recently proposed through the multicarrier transmission scheme of AMQD (adaptive multicarrier quadrature division) [2]. The AMQD allows improved secret key rates and higher tolerable excess noise in comparison with standard (referred to as single-carrier throughout) CVQKD. The multicarrier transmission granulates the information into several Gaussian subcarrier CVs, which are then transmitted through the Gaussian sub-channels. Particularly, the AMQD divides the physical Gaussian channel into several Gaussian sub-channels; each sub-channel is dedicated for the conveying of a given Gaussian subcarrier CV. The physical medium of the individual subcarriers are coherent quantum states, similar to single-carrier CVQKD. A multicarrier CVQKD also provides an unconditional security against the most powerful attacks [2, 4].

The proposed diversity extraction uses a sophisticated phase space constellation for the Gaussian sub-channels [4] which provides a natural framework to exploit the diversity patterns of the sub-channel transmittance coefficients. The diversity extraction can be applied for an arbitrary distribution of the sub-channel transmittance coefficients and, by exploiting some properties of the phase space constellation it does not require the use of a statistical model. The proposed phase space constellation offers an analogous criterion to an averaging over the statistics of the sub-channel transmittance coefficients. We compare the achievable performance of diversity extraction of multicarrier and single-carrier CVQKD.

This paper is organized as follows. Section 2 summarizes some preliminary findings. Section 3 defines the diversity space for CVQKD. Section 4 proposes the error analysis of diversity extraction of multicarrier CVQKD. Finally, Section 5 concludes the results.

## II. Multicarrier CVQKD

In this section we very briefly summarize the basic notations of AMQD from [2]. The following description assumes a single user, and the use of $n$ Gaussian sub-channels $\mathcal{N}_i$ for the transmission of the subcarriers, from which only $l$ sub-channels will carry valuable information.

In the single-carrier modulation scheme, the $j$-th input single-carrier state $|\varphi_j\rangle = |x_j + ip_j\rangle$ is a Gaussian state in the phase space $\mathcal{S}$, with i.i.d. Gaussian random position and momentum quadratures $x_j \in \mathbb{N}(0, \sigma_{\omega_0}^2)$, $p_j \in \mathbb{N}(0, \sigma_{\omega_0}^2)$, where $\sigma_{\omega_0}^2$ is the modulation variance of the quadratures. In the multicarrier scenario, the information is carried by Gaussian subcarrier CVs, $|\phi_i\rangle = |x_i + ip_i\rangle$, $x_i \in \mathbb{N}(0, \sigma_{\omega}^2)$, $p_i \in \mathbb{N}(0, \sigma_{\omega}^2)$, where $\sigma_{\omega}^2$ is the modulation variance of the subcarrier quadratures, which are transmitted through a noisy Gaussian sub-channel $\mathcal{N}_i$. Precisely, each $\mathcal{N}_i$ Gaussian sub-channel is dedicated for the transmission of one Gaussian subcarrier CV from the $n$ subcarrier CVs. (*Note*: index $l$ refers to the subcarriers, while index $j$, to the single-carriers, throughout the manuscript.) The single-carrier state $|\varphi_j\rangle$ in the phase space $\mathcal{S}$ can be modeled as a zero-mean, circular symmetric complex Gaussian random variable $z_j \in \mathcal{CN}(0, \sigma_{\omega_{z_j}}^2)$, with variance $\sigma_{\omega_{z_j}}^2 = \mathbb{E}\left[|z_j|^2\right]$, and with i.i.d. real and imaginary zero-mean Gaussian random components $\operatorname{Re}(z_j) \in \mathbb{N}(0, \sigma_{\omega_0}^2)$, $\operatorname{Im}(z_j) \in \mathbb{N}(0, \sigma_{\omega_0}^2)$.

In the multicarrier CVQKD scenario, let $n$ be the number of Alice's input single-carrier Gaussian states. Precisely, the $n$ input coherent states are modeled by an $n$-dimensional, zero-mean, circular symmetric complex random Gaussian vector

$$\mathbf{z} = \mathbf{x} + i\mathbf{p} = (z_1, \ldots, z_n)^T \in \mathcal{CN}(0, \mathbf{K_z}), \quad (1)$$

where each $z_j$ is a zero-mean, circular symmetric complex Gaussian random variable

$$z_j \in \mathcal{CN}(0, \sigma_{\omega_{z_j}}^2), \quad z_j = x_j + ip_j. \quad (2)$$

Specifically, the real and imaginary variables (i.e., the position and momentum quadratures) formulate $n$-dimensional real Gaussian random vectors, $\mathbf{x} = (x_1, \ldots, x_n)^T$ and $\mathbf{p} = (p_1, \ldots, p_n)^T$, with zero-mean Gaussian random variables. The Fourier transformation $F(\cdot)$ of the $n$-dimensional Gaussian random vector $\mathbf{v} = (v_1, \ldots, v_n)^T$ results in the $n$-dimensional Gaussian random vector $\mathbf{m} = (m_1, \ldots, m_n)^T$, precisely:

$$\mathbf{m} = F(\mathbf{v}) = e^{\frac{-\mathbf{m}^T \mathbf{A} \mathbf{A}^T \mathbf{m}}{2}} = e^{\frac{-\sigma_{\omega_0}^2 (m_1^2 + \ldots + m_n^2)}{2}}. \quad (3)$$

In the first step of AMQD, Alice applies the inverse FFT (fast Fourier transform) operation to vector $\mathbf{z}$ (see (1)), which results in an $n$-dimensional zero-mean, circular symmetric complex Gaussian random vector $\mathbf{d}$, $\mathbf{d} \in \mathcal{CN}(0, \mathbf{K_d})$, $\mathbf{d} = (d_1, \ldots, d_n)^T$. The $\mathbf{T}(\mathcal{N})$ transmittance vector of $\mathcal{N}$ in the multicarrier transmission is

$$\mathbf{T}(\mathcal{N}) = [T_1(\mathcal{N}_1), \ldots, T_n(\mathcal{N}_n)]^T \in \mathcal{C}^n, \quad (4)$$

where

$$T_i(\mathcal{N}_i) = \operatorname{Re}(T_i(\mathcal{N}_i)) + i\operatorname{Im}(T_i(\mathcal{N}_i)) \in \mathcal{C}, \quad (5)$$

is a complex variable, which quantifies the position and momentum quadrature transmission (i.e., gain) of the $i$-th Gaussian sub-channel $\mathcal{N}_i$, in the phase space $\mathcal{S}$, with real and imaginary parts

$$0 \leq \operatorname{Re} T_i(\mathcal{N}_i) \leq 1/\sqrt{2}, \text{ and } 0 \leq \operatorname{Im} T_i(\mathcal{N}_i) \leq 1/\sqrt{2}. \quad (6)$$

The Fourier-transformed transmittance of the $i$-th sub-channel $\mathcal{N}_i$ (resulted from CVQFT (continuous-variable quantum Fourier transform) operation at Bob) is denoted by

$$\left|F(T_i(\mathcal{N}_i))\right|^2. \quad (7)$$

The $n$-dimensional zero-mean, circular symmetric complex Gaussian noise vector $\Delta \in \mathcal{CN}(0, \sigma_\Delta^2)_n$ of the quantum channel $\mathcal{N}$, is evaluated as

$$\Delta = (\Delta_1, \ldots, \Delta_n)^T \in \mathcal{CN}(0, \mathbf{K}_\Delta), \quad (8)$$

where

$$\mathbf{K}_\Delta = \mathbb{E}\left[\Delta \Delta^\dagger\right], \quad (9)$$

with independent, zero-mean Gaussian random components

$$\Delta_{x_i} \in \mathbb{N}(0, \sigma_{\mathcal{N}_i}^2), \text{ and } \Delta_{p_i} \in \mathbb{N}(0, \sigma_{\mathcal{N}_i}^2), \quad (10)$$

with variance $\sigma_{\mathcal{N}_i}^2$, for each $\Delta_i$ of a Gaussian sub-channel $\mathcal{N}_i$, which identifies the Gaussian noise of the $i$-th sub-channel $\mathcal{N}_i$ on the quadrature components in the phase space $\mathcal{S}$.

The general model of AMQD is depicted in Fig. 1.

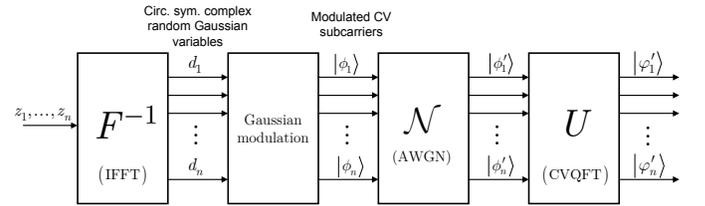

**Fig. 1.** The AMQD modulation scheme [2]. Alice draws an $n$-dimensional, zero-mean, circular symmetric complex Gaussian random vector $\mathbf{z}$, which are then inverse Fourier-transformed by $F^{-1}$. The resulting vector $\mathbf{d}$ encodes the subcarrier quadratures for the Gaussian modulation. In the decoding, Bob applies the $U$ unitary CVQFT on the $n$ subcarriers to recover the noisy version of Alice's original variable as a continuous variable in the phase space (IFFT – inverse fast Fourier transform, AWGN – additive white Gaussian noise, CVQFT – inverse continuous-variable quantum Fourier transform).

## III. DIVERSITY SPACE OF MULTICARRIER CVQKD

In a multicarrier CVQKD scenario, the term *diversity* is interpreted as follows. Let the $i$-th component $p_{j,i}$ of a given private random codeword $\mathbf{p}_j = (p_{j,1},\ldots,p_{j,l})^T$ to be transmitted through $\mathcal{N}_i$, where each Gaussian sub-channel is characterized by an independent transmittance coefficient $|(T_i(\mathcal{N}_i))|^2$. As a first approach, the number $l$ of the Gaussian sub-channels is identified as the *diversity* of $\mathcal{N}$. Specifically, the transmission can be utilized by a permutation phase space constellation $\mathcal{C}_{\mathcal{S}}^P(\mathcal{N})$ [4]. Using $P_i$, $i=2,\ldots,l$ random permutation operators, $\mathcal{C}_{\mathcal{S}}^P(\mathcal{N})$ can be defined for the multicarrier transmission as

$$\mathcal{C}_{\mathcal{S}}^P(\mathcal{N}) = (\mathcal{C}_{\mathcal{S}}(\mathcal{N}_1),\ldots,\mathcal{C}_{\mathcal{S}}(\mathcal{N}_l))$$
$$= (\mathcal{C}_{\mathcal{S}}(\mathcal{N}_1), P_2\mathcal{C}_{\mathcal{S}}(\mathcal{N}_1),\ldots, P_l\mathcal{C}_{\mathcal{S}}(\mathcal{N}_1)), \quad (11)$$

where $d_{\mathcal{C}_{\mathcal{S}}(\mathcal{N}_i)} = d_{\mathcal{C}_{\mathcal{S}}(\mathcal{N}_j)}$ is the cardinality of $\mathcal{C}_{\mathcal{S}}(\mathcal{N}_i)$. Using $\mathcal{C}_{\mathcal{S}}^P(\mathcal{N})$, the available degrees of freedom in the Gaussian link can be utilized, and the random permutation operators inject correlation between the $\mathcal{N}_i$ sub-channels via $P_i\mathcal{C}_{\mathcal{S}}(\mathcal{N}_1)$. In particular, for each Gaussian sub-channel, the distance between the phase space constellation points is evaluated by $\delta_i$, the normalized difference function. Assuming two $l$-dimensional input random private codewords $\mathbf{p}_A = (p_{A,1},\ldots p_{A,l})^T$ and $\mathbf{p}_B = (p_{B,1},\ldots p_{B,l})^T$ and two Gaussian sub-channels $\mathcal{N}_i$ and $\mathcal{N}_j$, $\delta_i$ is calculated precisely as follows:

$$\delta_i = \frac{1}{\sqrt{\frac{\sigma_{\omega'}^2}{\sigma_{\mathcal{N}^*}^2}}}(p_{A,i} - p_{B,i}), \quad (12)$$

where $\sigma_{\omega'}^2$ is the variance of the subcarriers, while $\sigma_{\mathcal{N}^*}^2$ is the noise variance of the sub-channels, respectively.

Particularly, for the $l$ Gaussian sub-channels the *product distance* $|\delta_{1\ldots l}|$ is as [18-20]

$$|\delta_{1\ldots l}|^2 > \left(c\frac{1}{12^{S(\mathcal{N}_i)}}\right)^l, \quad (13)$$

where $c>0$ is a constant. The maximization of this term ensures the maximization of the extractable diversity, and determines the $\tilde{p}_{err}$ pairwise worst-case error probabilities of $\mathbf{p}_A, \mathbf{p}_B$.

## IV. ERROR ANALYSIS

By using $\mathcal{C}_{\mathcal{S}}^P(\mathcal{N})$ and (13), the $\tilde{p}_{err}$ worst-case pairwise error probability can be decreased to the theoretical lower bound. We further reveal that in a multiuser CVQKD scenario, this condition can be extended simultaneously for all users. Let us assume that the $S'_k(\mathcal{N})$ secret key rate of user $U_k$, for $\forall k$, is fixed precisely as follows:

$$S'_k(\mathcal{N}) = \tfrac{\varsigma_k}{n_{\min}} P'(\mathcal{N}), \quad (14)$$

where $\varsigma_k > 0$ is referred to as the *degree of freedom ratio* of $U_k$, and $n_{\min} = \min(K_{in}, K_{out})$, where $K_{in}$ and $K_{out}$ refer to the number of sender and receiver users. As one can immediately conclude from (14), $S'_k(\mathcal{N}) \ll P'(\mathcal{N})$. Without loss of generality, for a given Gaussian sub-channel $\mathcal{N}_i$, we redefine $S'_k(\mathcal{N}_i)$, $\varsigma_{k,i} > 0$ precisely as

$$S'_k(\mathcal{N}_i) = \tfrac{\varsigma_{k,i}}{n_{\min}} P'(\mathcal{N}_i). \quad (15)$$

(*Note*: From this point, we use the complex domain formulas throughout the manuscript and $S'_k(\mathcal{N})$ and $S'_k(\mathcal{N}_i)$ are fixed to (14) and (15).). For a given $\mathcal{N}_i$, an $\mathrm{E}_{err}$ error event [18–20] is identified as follows:

$$\mathrm{E}_{err} \equiv \log_2\left(1 + |F(T_i(\mathcal{N}_i))|^2 (\mathrm{SNR}'_i)^*\right) < S'(\mathcal{N}_i), \quad (16)$$

and the probability of $\mathrm{E}_{err}$ at a given $S'(\mathcal{N}_i)$ is identified by the $p_{err}$ error probability as follows:

$$\mathrm{E}_{err} = p_{err}(S'_k(\mathcal{N}_i))$$
$$= \Pr\left(\begin{array}{c}\log_2\left(1 + |F(T_i(\mathcal{N}_i))|^2 (\mathrm{SNR}'_i)^*\right) \\ < S'(\mathcal{N}_i)\end{array}\right). \quad (17)$$

Particularly, by some fundamental argumentations on the statistical properties of a Gaussian random distribution [18–20], for $|F(T_i(\mathcal{N}_i))|^2 (\mathrm{SNR}'_i)^* \to 0$, $p_{err}(S'_k(\mathcal{N}_i))$ can be expressed as

$$p_{err}(S'_k(\mathcal{N}_i)) = \Pr\left(\begin{array}{c}|F(T_i(\mathcal{N}_i))|^2 (\mathrm{SNR}'_i)^* \log_2 e \\ < S'(\mathcal{N}_i)\end{array}\right), \quad (18)$$

while for $|F(T_i(\mathcal{N}_i))|^2 (\mathrm{SNR}'_i)^* \to \infty$, the corresponding error probability is as

$$p_{err}(S'_k(\mathcal{N}_i)) = \Pr\left(\begin{array}{c}\log_2\left(|F(T_i(\mathcal{N}_i))|^2 (\mathrm{SNR}'_i)^*\right) \\ < S'(\mathcal{N}_i)\end{array}\right). \quad (19)$$

Let $l=1$, that is, let's consider a single-carrier CVQKD, with $|F(T(\mathcal{N}))|^2$ of $\mathcal{N}$, with a secret key rate $S'(\mathcal{N})$. In this setting, $p_{err}$ is expressed precisely as [18]

$$p_{err}^{single}(S'_k(\mathcal{N})) = \Pr\left(\begin{array}{c}\log_2\left(1 + |F(T(\mathcal{N}))|^2 (\mathrm{SNR}')^*\right) \\ < S'(\mathcal{N})\end{array}\right)$$
$$= \Pr\left(|F(T(\mathcal{N}))|^2 < \tfrac{1}{(\mathrm{SNR}')^*}\right)$$
$$= \tfrac{1}{(\mathrm{SNR}')^*}, \quad (20)$$

by theory. Specifically, assuming a multicarrier CVQKD scenario with $l$ Gaussian sub-channels and secret key rate $S'(\mathcal{N}_i)$ per $\mathcal{N}_i$, $p_{err}^{AMQD}$ is derived as follows. Without loss of generality, we construct the set $\mathcal{T}$, such that

$$\mathcal{T}: \min_{\forall i}\left\{\left|F(T_i(\mathcal{N}_i))\right|\right\}, \quad (21)$$

where for $\forall i, i = 1,\ldots l$ the following condition holds:

$$F(T_i(\mathcal{N}_i)) \geq \tfrac{1}{(\text{SNR}')^*}. \quad (22)$$

In particular, the transmission through the Gaussian sub-channels is evaluated via set $\mathcal{T}$, which refers to the worst-case scenario at which a $S'(\mathcal{N}) > 0$ nonzero secret key rate is possible, by convention. Particularly, in (13), a given $\partial_i$ identifies the minimum distance between the normalized $2^{S'_k(\mathcal{N}_i)}$ points for the phase space constellation $\mathcal{C}'_\mathcal{S}(\mathcal{N}_i)$ of $\mathcal{N}_i$. Precisely, by fundamental theory [18], it can be proven that for an arbitrary distribution of the $F(T_i(\mathcal{N}_i))$ Fourier transformed transmittance coefficient, the maximized product distance function of (13) can be derived by an averaging over the following statistic $\mathcal{S}$:

$$\mathcal{S}: F(T_i(\mathcal{N}_i)) \in \mathcal{CN}\left(0, \sigma^2_{F(T_i(\mathcal{N}_i))}\right), \quad (23)$$

where $\sigma^2_{F(T_i(\mathcal{N}_i))} = \mathbb{E}\left[\left|F(T_i(\mathcal{N}_i))\right|^2\right]$, and $F(T_i(\mathcal{N}_i))$ is a zero-mean, circular symmetric complex Gaussian random variable with i.i.d. $\mathbb{N}\left(0, 0.5\sigma^2_{F(T_i(\mathcal{N}_i))}\right)$ zero-mean Gaussian random variables per quadrature components $x_i$ and $p_i$, for the $i$-th Gaussian subcarrier CV.

Putting the pieces together, the maximized product distance function $|\delta_{1...l}|$ of (13) precisely can be obtained via an averaging over the $\mathcal{S}$ statistics of (23); however, (23) is, in fact, strictly provides an analogous criteria of the worst-case $\tilde{p}_{err}$ situation in (21) via a sophisticated phase space constellation $\mathcal{C}_\mathcal{S}$, by theory [18–20]. In other words, set $\mathcal{T}$, as it is given in (21) together with $\mathcal{C}_\mathcal{S}$ represents a universal criteria and provides us an alternative solution to find the worst-case $\tilde{p}_{err}$ error probability for arbitrary distributed $F(T_i(\mathcal{N}_i))$ coefficients in a multicarrier CVQKD scenario. Specifically, some of these argumentations can be further exploited in our analysis.

First of all, by using (23), the averaged term $\tfrac{1}{l}\sum_l|F(T_i(\mathcal{N}_i))|^2$ can be modeled as a sum of $\mathcal{CN}\left(0, \sigma^2_{F(T_i(\mathcal{N}_i))}\right)$ distributed random variables, with zero mean and variance of $\sigma^2_{F(T_i(\mathcal{N}_i))}$ for each $\mathcal{N}_i$ sub-channels. Then, since $\tfrac{1}{l}\sum_l|F(T_i(\mathcal{N}_i))|^2$ is the averaged sum of $2l$ independent real Gaussian random variables, the distribution of $\tfrac{1}{l}\sum_l|F(T_i(\mathcal{N}_i))|^2$ precisely can be approximated by a $\chi^2_{2l}$ chi-square distribution with $2l$ degrees of freedom, by a density function $f(\cdot)$:

$$f(x) = \tfrac{1}{(l-1)!}x^{l-1}e^{-x}, \quad (24)$$

where $x \geq 0$. In particular, for $x \to 0$, the density can be written as

$$f(x) \approx \tfrac{1}{(l-1)!}x^{l-1}. \quad (25)$$

Thus, we arrive at $p_{err}^{AMQD}$ as

$$\begin{aligned}
p_{err}^{AMQD} &= \Pr\left(\tfrac{1}{l}\sum_l|F(T_i(\mathcal{N}_i))|^2 < \tfrac{1}{(\text{SNR}')^*}\right) \\
&= \int_0^{\tfrac{1}{(\text{SNR}')^*}} \tfrac{1}{(l-1)!}x^{l-1}dx \\
&= \tfrac{1}{l!}\tfrac{1}{\left((\text{SNR}')^*\right)^l} \\
&\approx \tfrac{1}{\left((\text{SNR}')^*\right)^l},
\end{aligned} \quad (26)$$

where the term $\tfrac{1}{l!}$ is negligible. Specifically, from (20) and (26), the $\delta$ *diversity parameter* picks up the following value in the single-carrier CVQKD setting:

$$\delta_{single} = 1, \quad (27)$$

while in the multicarrier CVQKD setting,

$$\delta_{AMQD} = l. \quad (28)$$

The result in (28) significantly depends on the properties the corresponding phase space constellation $\mathcal{C}_\mathcal{S}(\mathcal{N})$. From (31) it clearly follows that the extractable diversity $\delta$ determines the error probability of the transmission, and for higher $\delta$, the reliability of the transmission improves.

Particularly, in a *multiple-access* CVQKD scenario, there exists another degree of freedom in the channel, the number of information carriers allocated to a given user $U$. This type of degree of freedom is denoted by $\varsigma$ and is referred to as the *degree of freedom ratio*. Without loss of generality, in the function of $\varsigma > 0$ (27) and (28) precisely can be rewritten as

$$\delta_{single} = 1 - \varsigma, \quad (29)$$

while, in the multicarrier CVQKD setting, it refers to the ratio of the subcarriers allocated to a given user,

$$\delta_{AMQD} = l(1-\varsigma). \quad (30)$$

Thus, in a multicarrier CVQKD scenario with $l$ Gaussian sub-channels, for a given $\varsigma > 0$, the overall gain is $l$. As follows, using (29) and (30), the error probabilities can be rewritten precisely as

$$p_{err}^{single} = \tfrac{1}{\left((\text{SNR}')^*\right)^{\delta_{single}}} = \tfrac{1}{\left((\text{SNR}')^*\right)^{(1-\varsigma)}}, \quad (31)$$

$$p_{err}^{AMQD} = \tfrac{1}{\left((\text{SNR}')^*\right)^{\delta_{AMQD}}} = \tfrac{1}{\left((\text{SNR}')^*\right)^{l(1-\varsigma)}}. \quad (32)$$

The $p_{err}^{single}$ and $p_{err}^{AMQD}$ error probabilities of (31) and (32) for $(\text{SNR}')^* \geq 1$, for $l = 5, 10$ Gaussian sub-channels, and at $\varsigma = 0.6$ are compared in Fig. 2.

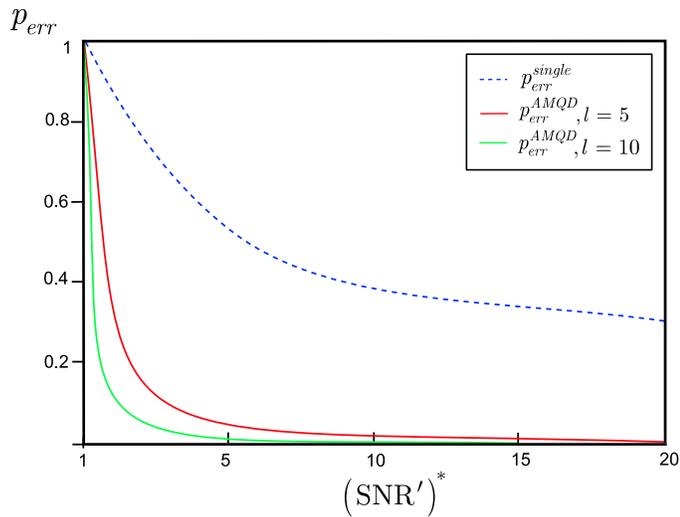

**Fig. 2.** The error probabilities in the single-carrier and multicarrier CVQKD (belonging to $l$ subcarriers), $(\text{SNR}')^* \geq 1$, $l = 5, 10$, and $\varsigma = 0.6$.

## V. CONCLUSION

The additional degree of freedom injected by the multicarrier transmission represents a significant resource to achieve performance improvements in CVQKD protocols. The proposed diversity extraction exploits those extra resources brought in by the multicarrier CVQKD modulation and is unavailable in a single-carrier CVQKD scheme. The results confirm that the possibilities in a multicarrier CVQKD significantly exceed the single-carrier CVQKD scenario. The available and efficiently exploitable extra resources have a crucial significance in experimental CVQKD, particularly in long-distance scenarios.

ACKNOWLEDGMENT

This work was partially supported by the GOP-1.1.1-11-2012-0092 (Secure quantum key distribution between two units on optical fiber network) project sponsored by the EU and European Structural Fund, by the Hungarian Scientific Research Fund - OTKA K-112125, and by the COST Action MP1006.